# Magneto-optical and magnetotransport properties of heavily Mn-doped GaMnAs

Shinobu Ohya

*Department of Electronic Engineering, The University of Tokyo, 7-3-1 Hongo, Bunkyo-ku, Tokyo 113-8656, Japan, and PRESTO Japan Science and Technology Agency, 4-1-8 Honcho, Kawaguchi, Saitama 332-0012, Japan*

Kenichi Ohno

*Department of Electronic Engineering, The University of Tokyo, 7-3-1 Hongo, Bunkyo-ku, Tokyo 113-8656, Japan*

Masaaki Tanaka

*Department of Electronic Engineering, The University of Tokyo, 7-3-1 Hongo, Bunkyo-ku, Tokyo 113-8656, Japan, and SORST Japan Science and Technology Agency, 4-1-8 Honcho, Kawaguchi, Saitama 332-0012, Japan*

We have studied the magneto-optical and magnetotrasnport properties of $Ga_{1-x}Mn_xAs$ thin films with high Mn concentrations ($x$= 12.2 - 21.3%) grown by molecular-beam epitaxy. These heavily Mn-doped GaMnAs films were formed by decreasing the growth temperature to as low as 150-190ºC and by reducing the film thickness to 10 nm in order to prevent precipitation of hexagonal MnAs clusters. Magnetic circular dichroism (MCD) and anomalous Hall effect measurements indicate that these GaMnAs films have the nature of intrinsic ferromagnetic semiconductors with high ferromagnetic transition temperature up to 170 K.

III-V-based ferromagnetic semiconductor (FMS) GaMnAs[1,2] is a good model system for future spintronic devices. Recently, large tunneling magnetoresistance (TMR)[3-5] and TMR oscillation induced by the resonant tunneling effect[6] were observed in GaMnAs-based single-barrier and double-barrier heterostructures, respectively. These results indicate that the FMS heterostructures are very promising for future quantum devices using spin degrees of freedom. For realizing such devices operating at room temperature, however, it is important to increase the Curie temperature ($T_C$) of $Ga_{1-x}Mn_xAs$, which is limited nowadays to 173 K (at $x$=8%).[7] The mean-field Zener model[8] of carrier-mediated ferromagnetism predicts that $T_C$ of $Ga_{1-x}Mn_xAs$ is proportional to $xp^{1/3}$, where $p$ is the hole concentration and $x$ is the Mn concentration, thus higher $T_C$ values are expected in GaMnAs with a higher Mn concentration if sufficiently high hole concentration is obtained. In fact, in the Mn δ-doped AlGaAs/GaAs modulation-doped heterostructure where a high Mn content is locally achieved within a few MLs, high $T_C$ values of 192-250 K were reported.[9] Meanwhile, the growth of $Ga_{1-x}Mn_xAs$ alloy films with $x$>10% has been very difficult because precipitation of hexagonal MnAs clusters more easily occurs during growth with higher $x$. Until now, no papers have been published on $Ga_{1-x}Mn_xAs$ alloys with $x>10\%$.

In this letter, we report the successful growth and properties of $Ga_{1-x}Mn_xAs$-alloy films with high $x$ ranging from 12.2 to 21.3%. In order to prevent the precipitation during growth, samples were grown by decreasing the growth temperature $T_S$ to as low as 150-190°C and by reducing the film thickness $d$ to 10 nm.[10] We measured the magneto-optical and magnetotransport properties, indicating that these GaMnAs films with high $x$ have intrinsic FMS features like GaMnAs with low $x$ (<10%). Furthermore, we found that post-growth annealing is very effective to increase $T_C$ of these heavily Mn-doped GaMnAs, and high $T_C$ values up to 170 K were obtained by annealing these samples.

The heavily Mn-doped $Ga_{1-x}Mn_xAs$ films with $x$=12.2, 15.2, 17.6, and 21.3% were grown by low-temperature molecular-beam epitaxy (LT-MBE) with the growth conditions shown in Table I. The growth procedure was as follows. We grew a GaAs buffer layer on a semi-insulating GaAs(001) substrate at 600°C, and decreased $T_S$ to 150-190°C. After that, a 10-nm thick $Ga_{1-x}Mn_xAs$ thin film was grown on this buffer layer. In the cases of $x$=17.6% and 21.3%, a 1-nm thick GaAs-cap layer was grown on the top of GaMnAs layer in order to prevent the formation of the MnAs clusters by segregated Mn atoms on the surface after the growth. The Mn concentration $x$ was determined by the relationship between the Mn flux and Mn cell's temperature, and the Mn flux data were precisely estimated by measuring the thickness of epitaxial MnAs films grown in our MBE system. Figure 1 shows the reflection high-energy electron diffraction (RHEED) pattern during the growth of $Ga_{0.824}Mn_{0.176}As$. The RHEED pattern was streaky 1×1 or 1×2, indicating two-dimensional growth of single-crystal films without phase decomposition or precipitation.[2] After the MBE growth, these samples were annealed in air at $T_A$=140-160°C for long hours ($t_A$~100 h), and their post-growth annealing conditions are also described in Table I. We measured the $T_C$ values of these samples before and after annealing with various $t_A$ (0-350 h), and confirmed that $T_C$ completely saturates at the annealing times shown in Table I in all the samples.

We carried out magnetic circular dichroism (MCD) measurements in a reflection setup. Here, the MCD intensity is expressed as

$$\text{MCD}[\text{deg}] = \frac{90}{\pi}\frac{R_+ - R_-}{R_+ + R_-} \propto \Delta E \frac{1}{R}\frac{dR}{dE}, \quad (1)$$

where $R_+$ and $R_-$ are optical reflectance for σ+ and σ- circularly polarized light, respectively, $\Delta E$ is the Zeeman splitting energy, $R$ is the total reflectance, and $E$ is the photon energy. Since MCD is proportional to $dR/dE$, the MCD spectrum is strongly associated with the band structure of the measured material. From the shape of the MCD spectrum and its magnetic field dependence, we can distinguish whether the measured magnetization comes from FMS or from magnetic precipitations. Since MCD is directly proportional to the Zeeman splitting $\Delta E$ which is proportional to vertical component of magnetization $M$, the magnetization characteristics (corresponding to $M$-$H$ curves) can be obtained from the magnetic field dependence of the MCD intensity. The existence of the Zeeman splitting of the host semiconductor-like bandstructure is one of the most essential features of FMS. Thus, by carrying out careful MCD characterizations, we can obtain the direct evidence for 'intrinsic'

FMS, where the ferromagnetism comes from the FMS not from magnetic precipitations.[11]

Figure 2(a) shows the MCD spectra of annealed $Ga_{0.878}Mn_{0.122}As$ measured at 11 K with the magnetic fields of 1 kOe (dotted black curve), 3 kOe (thin grey curve), and 10 kOe (thick black curve) applied perpendicular to the film plane. As a reference, the MCD spectrum of $Ga_{1-x}Mn_xAs$ with low $x$ of 0.074 at 5 K is shown in Fig. 2(c).[12] In all the spectra of Fig. 2(a), two broad negative peaks at around 1.9-2.0 eV and 3 eV are seen, corresponding to the optical transitions at Γ and Λ critical points, respectively. These spectral shapes are very similar to that obtained in GaMnAs with low $x$ shown in Fig. 2(c), indicating that the band structure of annealed $Ga_{0.878}Mn_{0.122}As$ is of zinc-blende type, just as GaMnAs with low $x$. On the other hand, the energy positions of these MCD peaks of heavily Mn-doped GaMnAs shown in Fig. 2(a) are shifted by 0.2-0.3 eV to higher energy from those of GaMnAs with low $x$. (Fig. 2(c)) We found that this blue shift occurs also by the post-growth annealing. In the MCD spectrum of the as-grown $Ga_{0.878}Mn_{0.122}As$ (not shown), MCD peaks were observed at 1.8 and 2.9 eV, which are lower than those of the annealed $Ga_{0.878}Mn_{0.122}As$. The blue shift at around the $E_0$ energy is probably partly due to the change of the carrier concentration, *i.e.* by the Moss-Burstein shift induced by the increase of the Fermi energy, which is higher than the valence band in terms of hole energy.[13] But the reason for the blue shift at around the $E_1$ energy is not clear at present (The band gap energy at Λ is probably modified by heavy Mn doping). Figure 2(b) shows MCD spectra with various magnetic fields shown in Fig. 2(a) when normalized by the intensity at 1.9 eV. The MCD spectra at any magnetic field can be superimposed on a single spectrum over the whole photon-energy range. This indicates that the MCD signal comes from the FMS phase of GaMnAs, not from other materials,[14] which means that the intrinsic FMS feature is maintained in GaMnAs even with such a high Mn concentration. We confirmed that this intrinsic FMS feature is maintained also in other GaMnAs samples with high $x$ (12.2%-21.3%) investigated in this study.

Figure 3 shows the magnetic-field dependence of the MCD intensity ($\propto M$) of the annealed $Ga_{0.878}Mn_{0.122}As$ with a photon energy of 3.0 eV at various temperatures of 30 K (blue curve), 90 K (green curve), 130 K (yellow curve), and 170 K (red curve) with a magnetic field applied perpendicular to the film plane. Clear ferromagnetic order with the in-plane easy magnetization axis was observed up to 170 K.

Figure 4 shows the temperature dependence of the resistivity of the heavily Mn-doped GaMnAs samples grown in this study. Thin curves are those of as-grown samples and thick curves are those of annealed samples. In all the GaMnAs samples with various Mn concentrations investigated here, resistivity was remarkably decreased after the post-growth annealing, meaning that the Mn interstitial ($Mn_I$) defects, which act as double donors and compensate holes, decreased by annealing. During the annealing process, the $Mn_I$ atoms are diffused to the free sample surface and are passivated by oxidation.[15 - 18] On the other hand, in both of the as-grown and annealed samples, there is a tendency that resistivity increases with increasing the Mn concentration $x$.

We carried out the anomalous Hall effect measurements for these GaMnAs samples with a van der Pauw method. (The Hall resistance $R_{Hall}$ vs. magnetic field $B$ data was obtained by extracting the odd-function component from the raw Hall resistance $R_{raw}$ vs. $B$ data in order to exclude the component of the even-function magnetoresistance.[19]) In FMS materials, $R_{Hall}$ can be approximated by $R_{Hall} \approx c\rho(B)^n M/d$ when temperature is below $T_C$, where $c$ is a constant, $\rho(B)$ is the resistivity, and $n$ is a constant related to the scattering mechanism. If $\rho(B)$ does not largely depend on $B$ in the range of $B$ used in the measurements, the $R_{Hall}$-$B$ curve should be equivalent to the *MCD*-$B$ curve, because both of $R_{Hall}$ and MCD is directly proportional to $M$. This is also a very important feature of FMS. The black solid curves shown in Fig. 3 are the $R_{Hall}$-$B$ curves of annealed $Ga_{0.878}Mn_{0.122}As$ at 30, 90, 130, and 170 K. The shapes of the $R_{Hall}$-$B$ curves (black) at all the temperatures are in almost perfect agreement with the MCD-$B$ curves (colors).

In order to determine the exact $T_C$ value of the annealed $Ga_{0.878}Mn_{0.122}As$, we carried out the Arrott plot ($R_{Hall}^2$ vs. $B/R_{Hall}$) with the anomalous Hall effect data. Note that $R_{Hall}$ is directly proportional to $M$ below $T_C$, thus the $R_{Hall}^2$ vs. $B/R_{Hall}$ plot corresponds to the conventional Arrott plot ($M^2$-$B/M$).[20] The Arrott plot is shown in the inset of Fig. 3, indicating that $T_C$ of this annealed $Ga_{0.878}Mn_{0.122}As$ sample is 170 K, which is close to the highest record $T_C$ of GaMnAs.[7]

In Table I, we show $T_C$ values of the as-grown and annealed GaMnAs samples obtained by the above mentioned method. The $T_C$ values of all the samples were significantly increased by the low-temperature annealing. However, $T_C$ does not increase but decreases with increasing $x$ against the prediction of the mean field theory, which means that the $Mn_I$ defect concentration of our samples increases with increasing $x$. This result indicates that the Mn atoms are more likely to be incorporated into the interstitial site than into the substitutional site with increasing $x$ (or decreasing $T_S$). The reason that $T_C$ does not increase with $x$ even in the annealed samples is probably because there is an upper limit of the quantity of the $Mn_I$ atoms which can be passivated by oxidation. For obtaining higher $T_C$ in such a high $x$ region, it is important to find a way to remove the $Mn_I$ defects more effectively.

In summary, we have successfully grown the GaMnAs films with a high Mn concentration from 12.2 to 21.3% by decreasing the growth temperature to 150-200ºC and by reducing the film thickness to 10 nm. MCD and anomalous Hall effect measurements of these films indicate that GaMnAs maintains an intrinsic FMS feature even in such a high Mn-concentration region. High $T_C$ values up to 170 K were observed in these samples annealed in air at 160ºC for ~100 h.

This work was partly supported by PRESTO/SORST of JST, Grant-in-Aids for Scientific Research, IT Program of RR2002 of MEXT, and Kurata-Memorial Hitachi Science & Technology Foundation.

TABLE I. Growth- and annealing conditions and the obtained $T_C$ values of the heavily Mn-doped $Ga_{1-x}Mn_xAs$ samples investigated in this study. $d$, $T_S$, $T_A$ and $t_A$ are the film thickness of the GaMnAs layer, the growth temperature, the annealing temperature, and the annealing time, respectively. These $T_C$ values were estimated by the anomalous Hall effect measurements. The description of "<20" in the $T_C$ item means that we could not estimate the exact $T_C$ value because the resistivity was too high to measure in our measurement setup.

| $x$ (%) | $Ga_{1-x}Mn_xAs$ | | $T_C$ [K] | | Anneal conditions | |
|---|---|---|---|---|---|---|
| | $d$ [nm] | $T_S$ [°C] | As grown | annealed | $T_A$[°C] | Time $t_A$ [h] |
| 12.2 | 10 | 190 | 77 | 170 | 160 | 123.5 |
| 15.2 | 10 | 168 | 69 | 169 | 160 | 95 |
| 17.6 | 10 | 167 | 20 | 144 | 140 | 114 |
| 21.3 | 10 | 152 | < 20 | 125 | 140 | 118 |

**Figure captions**

FIG. 1.  RHEED pattern observed during the growth of $Ga_{0.824}Mn_{0.176}As$ when the electron beam azimuth is [$\bar{1}$10].

FIG. 2.  (a) MCD spectra of annealed $Ga_{0.878}Mn_{0.122}As$ measured in a reflection set up at 11 K when magnetic fields of 1 kOe (dotted black curve), 3 kOe (thin grey curve), and 10 kOe (thick black curve) were applied perpendicular to the film plane.  (b) MCD spectra shown in (a) when normalized by the intensity at 1.9 eV.  (c) The MCD spectrum of $Ga_{1-x}Mn_xAs$ with low $x$ of 0.074 at 5 K.[12]

FIG. 3.  Magnetic field $B$ dependence of MCD (left axis, color curves) and of Hall resistance $R_{Hall}$ (right axis, black curves) of annealed $Ga_{0.878}Mn_{0.122}As$ measured at 30, 90, 130, and 170 K.  The magnetic field was applied perpendicular to film plane.  The MCD measurements were carried out with photon energy of 3.0 eV in a reflection setup.  The inset shows the Arrott plot ($R_{Hall}^2$ vs. $B/R_{Hall}$) obtained by the anomalous Hall effect data at 156, 165, 168, and 170 K.

FIG. 4.  Temperature dependence of resistivity of as-grown (thin curves) and annealed (thick curves) $Ga_{1-x}Mn_xAs$ with $x$ of 12.2, 15.2, 17.6, and 21.3%.  Note that the data of the annealed samples with $x$=12.2 and 15.2% are almost overlapped.

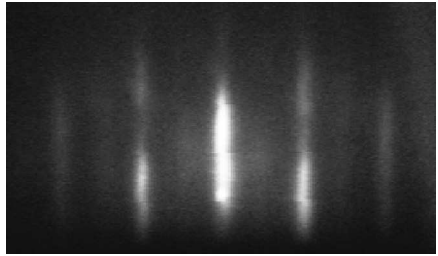

Figure 1

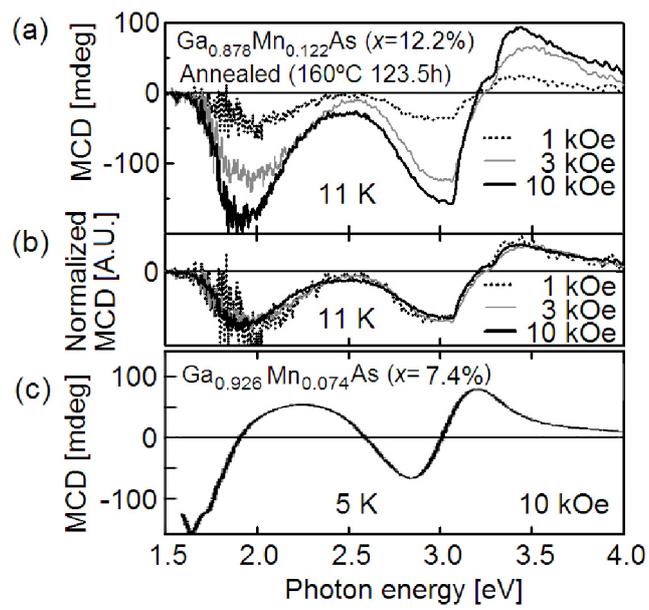

Figure 2

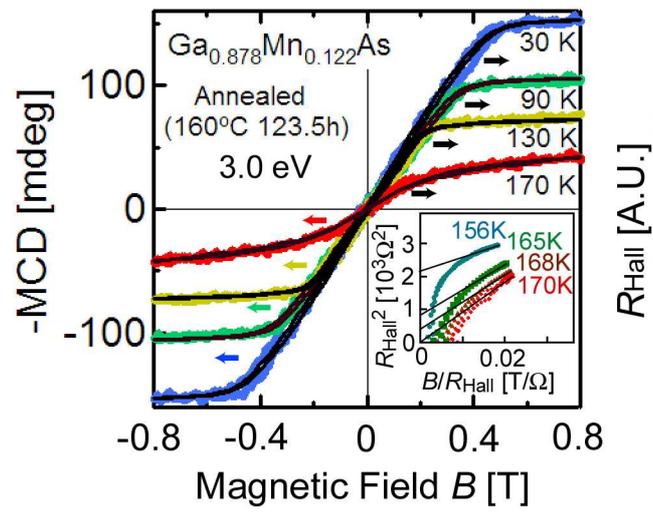

Figure 3

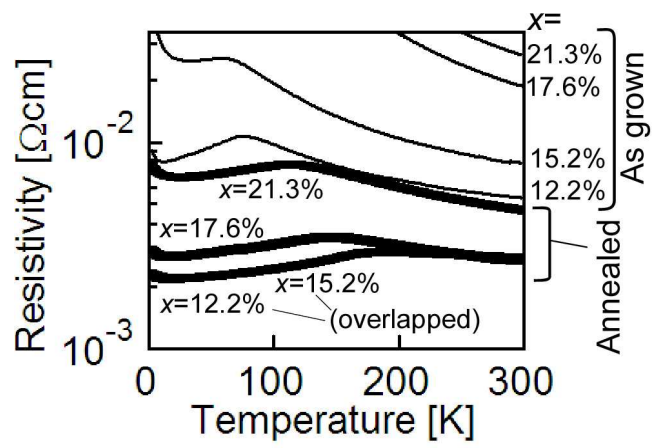

Figure 4